\documentstyle[sprocl]{article}

\def\Journal#1#2#3#4{{#1} {\bf #2}, #3 (#4)}

\begin{document}

\title{ROTATING INCOMPRESSIBLE PERFECT FLUID SOURCE OF THE NUT METRIC}

\author{ L\'{A}SZL\'{O} \'{A}. GERGELY$^{1,2}$, ZOLT\'{A}N
PERJ\'{E}S$^{1}$ and GYULA FODOR$^{1}$ }

\address{$^{1}$KFKI Research Institute for Particle and Nuclear Physics,\\ 
Budapest 114, P.O.Box 49, H-1525 Hungary\\
$^{2}$Laboratoire de Physique Th\'{e}orique, Universit\'{e} Louis Pasteur,\\
3-5 rue de l'Universit\'{e} 67084 Strasbourg Cedex, France}

\maketitle\abstracts{A global solution of the Einstein equations is
given, consisting of a perfect fluid interior and a vacuum 
exterior. The rigidly rotating and incompressible perfect fluid is 
matched along the hypersurface of vanishing pressure with the stationary 
part of the Taub-NUT metric. The fluid core generates a negative-mass 
NUT space-time. The matching procedure leaves 
one parameter in the global solution. }

\section{Introduction}
 
 Finding examples of space-times consisting of a rotating core and 
vacuum exterior is one of the many difficult issues in general relativity. A 
recent discussion on the matching of an axisymmetric rotating interior fluid
with ambient vacuum by Mars and Senovilla \cite{MaSe} reached the 
conclusion that the conditions of continuity of the first and 
second fundamental forms represents an overdetermined system. 
 
In this paper we construct a space-time containing a rotating perfect fluid 
surrounded by vacuum. We join a rigidly rotating incompressible perfect fluid 
solution \cite{F,Marklund,PFGM} with the Taub-NUT metric
\cite{Taub,NUT}, motivated by the isomorphic Killing algebras and by 
the peculiarities in the causal behavior of both space-times. What we get, 
however, is not what one would ideally expect since the fluid core does not 
rotate in the usual sense because it is locally rotationally symmetric (LRS).
 
The perfect fluid space-time has been described in detail elsewhere \cite{PFGM}. 
We cut this space-time along the hypersurface of zero pressure, which we make the 
common boundary with the vacuum region. We then apply the Darmois-Israel
matching procedure \cite{Darmois,Israel}. The continuity of both the
induced metric and extrinsic curvature lead to severe restrictions on the
parameters of the two solutions: all of them but one are frozen at
some specific values. This is still a fortunate result since there are more equations to fulfill than parameters.
 
Our signature convention is $\left( +---\right) $ and we choose Einstein's
gravitational constant $8\pi G=1.$
 
\section{Joining the two metrics}
 
The interior solution was found by Ferwagner \cite{F} and 
later rediscovered by Marklund \cite{Marklund}. In a recent work \cite{PFGM} 
it has been cast in a simple form with its properties further analyzed. 
This perfect fluid metric is:
\begin{eqnarray}
ds^{2}&=&\sin ^{4}\chi (dt+2R\cos \theta d\varphi )^{2}-2R\sin ^{2}\chi d\chi
(dt+2R\cos \theta d\varphi )\nonumber\\
&-&R^{2}\sin ^{2}\chi (d\theta ^{2}+\sin
^{2}\theta d\varphi ^{2})\ ,  \label{PFG}
\end{eqnarray}
where $R$ is the only parameter of the solution. The density is constant, $
\mu =6/R^{2}$, and the fluid is rigidly rotating with the four-velocity
\begin{equation}
u=\sin ^{-2}\chi \frac{\partial }{\partial t}\ .
\end{equation}
The pressure depends on the radial variable $\chi $ as
\begin{equation}
p=\frac{4}{R^{2}}\sin ^{-2}\chi -\frac{6}{R^{2}}\ .
\end{equation}
The condition of vanishing pressure selects a hypersurface $\chi=\chi_1$ 
defined by
\begin{equation}
\sin \chi _{1}=\sqrt{\frac{2}{3}.}  \label{chi1}
\end{equation}
At $\chi <\chi _{1}$ the pressure is positive, diverging in the origin,
where the metric is also singular. We cut the space-time along the embedding
hypersurface $\chi =\chi _{1}$ keeping the part where $\chi <\chi _{1}$.
 
The Taub-NUT metric has become renowned for being 'a counterexample to almost
anything' \cite{Misner}. The extension of the originally homogeneous Taub
metric \cite{Taub} is done by introducing new coordinates, obtaining a
space-time in which homogeneous and stationary parts are glued along Cauchy
horizons. In these coordinates the Taub-NUT metric is
\begin{eqnarray}
ds^{2}&=&\frac{r^{2}-2mr-l^{2}}{r^{2}+l^{2}}(d\tau+2l\cos \theta
d\varphi
)^{2}-2dr(d\tau+2l\cos \theta d\varphi )\nonumber\\
&-&\left( r^{2}+l^{2}\right)
(d\theta ^{2}+\sin ^{2}\theta d\varphi ^{2})\   \label{NUT}
\end{eqnarray}
with the parameters $m$ and $l$. The domain lying
between the roots $r_{\pm }=m\pm \sqrt{m^{2}+l^{2}}$ of the numerator of the
$g_{\tau\tau}$ metric coefficient describes a homogeneous cosmology. Outside the roots the metric (\ref{NUT}) is stationary.
 
The striking similarity in the appearence of the two metrics (\ref{PFG}) and
(\ref{NUT}) allows us to immediately list the conditions for continuity of
the induced metrics at the junction hypersurface $\chi =\chi _{1}$ and $r=r_{1}.$ These
conditions are:
\begin{eqnarray}
r_{1}^{2}+l^{2} &=&R^{2}\sin ^{2}\chi _{1}  \label{1a} \\
l &=&\alpha R  \label{1b} \\
\alpha ^{2}\frac{r_{1}^{2}-2mr_{1}-l^{2}}{r_{1}^{2}+l^{2}} &=&\sin ^{4}\chi
_{1}.  \label{1c}
\end{eqnarray}
At the junction the angles $\theta $ and $\varphi $ of the two metrics
are identified and the relation $\tau=\alpha t$ is assumed. Thus
altogether
there are four parameters in the global space-time: $\alpha ,m,l$ and $R$.
From (\ref{chi1}) and (\ref{1a}) the coordinate $r_1$ at the
junction is expressed in terms of the other parameters:
\begin{equation}
r_{1}=R\sqrt{\frac{2}{3}-\alpha ^{2}},  \label{r1}
\end{equation}
while (\ref{1c}) can be regarded as the defining equation for the parameter $
m$ in terms of the rest of the parameters.
 
Let us denote the space-time coordinates by $x^{a}=\left(
t,\rho ,\theta ,\varphi \right) $ where $\rho $ denotes the radial
coordinate, $\chi $ or $r$.
The coordinates $\xi ^{\alpha }=\left( t,\theta ,\varphi \right) $ can be
chosen as coordinates in both embeddings of the junction hypersurface. A holonomic
basis is thus given by
$e_{(\alpha)}^{a}=\partial x^{a}/\partial \xi^{\alpha}
=\delta _{\alpha }^{a}.$ The extrinsic curvature is
\begin{equation}
K_{\alpha \beta }=e_{(\alpha )}^{a}e_{(\beta )}^{b}\nabla _{a}n_{b}=-\Gamma
_{\alpha \beta }^{1}(-g^{11})^{-1/2}.
\end{equation}
The last equality applies because the normal to the junction hypersurface is
$n_{a}=(g^{11})^{-1/2}\delta _{a}^{1}$ for both metrics,
$g^{11}$ being $-1/R^2$ for
the fluid metric and $-\left( r^{2}-2mr-l^{2}\right) /\left(
r^{2}+l^{2}\right) $ for the Taub-NUT metric.
 
The condition to have no jump in the extrinsic curvature at the junction,
thus to have no distributional matter shell separating the two domains,
after some algebra and employing the relations (\ref{1b}), (\ref{1c}) and
(\ref{r1}), translates to:
\begin{equation}
\alpha =\frac{2}{3},\qquad
l=\frac{2}{3}R,\qquad
m=-\frac{2\sqrt{2}}{3}R.
\end{equation}
 
Thus all but one parameters are fixed, We end up with a
Universe which has a single parameter $R$. The mass parameter turns out 
to be negative.
 
The Cauchy horizons $r_{\pm }=2R\left( -\sqrt{2}+\sqrt{3}\right)
/3<r_{1}=R\sqrt{2}/3$ of
the Taub-NUT solution lie on the fluid side of  the junction hypersurface,
thus they are not present in the global space-time. Hence we have
a fluid interior joined with a stationary region of the Taub-NUT
space-time.
 
The causal behaviour of the global solution we have obtained is
of interest. In {\sl both} domains, at the junction hypersurface,
the vector  $e_{(\varphi )}^{a}$ becomes null at $\tan \theta _{\pm }=\pm 2
\sqrt{2/3}$ and timelike at $\theta \in [0,\theta _{+})\cup (\theta _{-},\pi
].$ This implies the existence of closed timelike curves, a familiar feature
of the Taub-NUT space-time.

The above discussion was generalized recently \cite{BFGMP} for a class of 
LRS perfect fluid space-times, all of them 
being sources for the NUT metric.
 
\section*{Acknowledgments}
 
This research has been supported by OTKA grants T17176, D23744 and T022563.
L.\'{A}.G was partially supported by the Hungarian State E\"{o}tv\"{o}s
Fellowship.


\section*{References}
\begin{thebibliography}{99}
\bibitem{MaSe}  M. Mars and J.M.M. Senovilla, 
                \Journal{{\em Mod. Phys. Lett.} A}{13}{1509}{1998}.
 
\bibitem{F} R. Ferwagner, in {\em Relativity Today}, ed. Z.Perj\'es (Nova Sci., 1992), p 133.
 
\bibitem{Marklund} M. Marklund, 
                \Journal{\em Class. Quant. Grav.}{14}{1267}{1997}.
 
\bibitem{PFGM}  Z. Perj\'{e}s, Gy. Fodor, L.\'{A}. Gergely and M.
                Marklund, gr-qc/9806095.
 
\bibitem{Taub} A. Taub, \Journal{\em Ann. Math.}{53}{472}{1951}.
 
\bibitem{NUT}  E.T Newman, L. Tamburino and T.J. Unti, 
               \Journal{\em J. Math. Phys.}{4}{915}{1963}.
 
\bibitem{Misner}  C.W. Misner in {\em Relativity Theory and Astrophysics I: 
Relativity and Cosmology, Lectures in Applied Mathematics, Volume 8
 }, ed. J. Ehlers (American Mathematical Society, 160-9, 1967).
 
\bibitem{Darmois}  G. Darmois in {\em M\'{e}morial de Sciences 
Math\'{e}matiques, Fascicule XXV, Chap. V.} (Gauthier-Villars, Paris, 1927).
 
\bibitem{Israel}  W. Israel, \Journal{{\em Nuovo Cimento} B}{XLIV}{4349}{1966}.

\bibitem{BFGMP}  M. Bradley, Gy. Fodor, L.\'{A}. Gergely, M. Marklund and 
Z. Perj\'{e}s, gr-qc/9807058.

\end{thebibliography}
\end{document}